# Toward Production-Ready Federated Learning in Healthcare: Privacy, Orchestration, and Governance in MLOps

Sakshi Gorkhali
Jarvis College of Computing and Digital Media
DePaul University
Chicago Illinois United States
sgorkhal@depaul.edu

Jonesh Shrestha
Jarvis College of Computing and Digital Media
DePaul University
Chicago Illinois United States
jshresth@depaul.edu

## ABSTRACT

Healthcare organizations often cannot freely centralize patient data because medical records are sensitive, regulated, and institutionally controlled. Federated learning offers a practical alternative by allowing hospitals and clinics to train a shared model while keeping raw data local. However, federated learning is not automatically production-ready or private by default. Model updates can still leak information, and decentralized training introduces operational challenges in deployment, monitoring, rollback, debugging, and governance. This paper examines how MLOps practices and the emerging idea of Federated Learning Operations (FLOps) can make federated healthcare machine learning systems scalable, reliable, and trustworthy. It answers three research questions: how containerization and orchestration support federated deployment, how privacy-preserving mechanisms affect trade-offs among privacy, utility, scalability, and operational complexity, and which post-deployment practices are most important for long-term governance. The central argument is that federated healthcare ML requires more than privacy-preserving algorithms. It needs an integrated MLOps architecture that combines reproducible deployment, secure orchestration, model versioning, audit logging, drift monitoring, heterogeneity management, and clear governance.

## CCS CONCEPTS

• Computing methodologies → Machine learning • Security and privacy → Privacy-preserving protocols • Software and its engineering → Software organization and properties → Software system structures → Software architectures • Applied computing → Health care information systems

## KEYWORDS

Federated learning, healthcare AI, MLOps, FLOps, privacy-preserving machine learning, secure aggregation, differential privacy, model governance, drift monitoring

## 1 Introduction

Healthcare AI depends on diverse data, but hospitals and research centers often cannot pool raw patient records into a single repository. HIPAA requires administrative, physical, and technical safeguards for electronic protected health information in the United States [1], while the GDPR treats health data as a special category subject to stricter protections in the European Union [2]. These constraints make centralized model training difficult in many healthcare settings.

Federated learning (FL) addresses this problem by moving training to the data instead of moving data to a central server. In a typical FL system, each hospital trains the model on its own local data and sends only model updates, such as weights or gradients, to a coordinating server. The server combines those updates into a new global model and sends the improved model back to the hospitals for the next round of training [3]. This allows institutions to collaborate without directly sharing raw patient records.

This approach is especially useful in healthcare, where model performance can improve when training reflects data from multiple institutions, patient populations, and clinical settings. Rieke et al. describe FL as a promising direction for digital health because it can help overcome data silos while respecting institutional boundaries [4]. However, FL does not remove all privacy and production risks. Model gradients or parameter updates may still leak information about local training data if additional protections are not used [5]. Therefore, FL should be understood as reducing data-centralization risk, not as a guarantee of privacy by itself.

This creates a direct MLOps problem. A production healthcare FL system must not only train a model, but also control versions, audit decisions, monitor drift, manage heterogeneous clients, handle failed participants, and support rollback when performance degrades. This paper addresses three research questions:

**RQ1:** How can a containerized and orchestrated MLOps architecture support scalable and reliable federated learning deployment across healthcare institutions while preserving data locality?

**RQ2:** How do privacy-preserving mechanisms such as secure aggregation, differential privacy, and encryption affect the trade-offs among patient privacy, model utility, scalability, and operational complexity?

**RQ3:** Which post-deployment MLOps practices are most important for ensuring trust, reproducibility, and long-term reliability in federated healthcare ML systems?

## 2 Literature Review

Federated learning was popularized by McMahan et al., who introduced Federated Averaging (FedAvg) [3]. The idea behind FedAvg is that each client starts with the same global model, trains it locally for a short time, and sends the updated model parameters back to the server. The server then averages the updates from participating clients to create the next global model. This process repeats across multiple rounds. FedAvg made FL practical by reducing the need to move raw data while still allowing many clients to contribute to shared learning.

Healthcare is a strong use case for FL because patient data is naturally distributed across hospitals, clinics, and research centers. Rieke et al. explain that FL can support digital health collaboration without requiring institutions to surrender control of their raw data [4]. This is important because healthcare data is not only private, but also shaped by local demographics, equipment, workflows, and clinical practices. A model trained at one hospital may not generalize well to another hospital, so collaboration can improve robustness. At the same time, this same diversity creates challenges because FL must handle non-identical data distributions across sites.

A common misconception is that FL is automatically private because raw data never leaves the local institution. Gradient inversion attacks, such as those demonstrated by Geiping et al., show that a malicious server or other attacker may sometimes reconstruct sensitive training examples or infer private information from client updates that are not protected by privacy-preserving mechanisms [5]. Three main techniques help address this risk: secure aggregation, which lets the server compute an aggregate without seeing individual updates [6]; differential privacy, which adds calibrated noise to limit what can be inferred about any individual [7]; and encryption-based approaches such as homomorphic encryption. Each involves different trade-offs between the level of privacy protection, model accuracy, and operational complexity.

MLOps is the set of practices for deploying, monitoring, and maintaining machine learning systems in production. Breck et al. show that production ML requires testing, monitoring infrastructure, canarying, and rollback capabilities to reduce operational risk [8]. These practices are even more important in FL because the model is trained and evaluated across multiple institutions instead of one centralized environment. If one hospital has outdated data, poor connectivity, or different preprocessing logic, the whole federated workflow can become unreliable.

The growing field of FLOps (Federated Learning Operations) extends MLOps principles to the federated lifecycle. Moon et al. propose a FedOps platform that uses Kubernetes and cloud-native tooling to manage FL lifecycle activities from integration and testing through deployment and monitoring [9]. Kukkaro et al. provide a systematic mapping study showing that while FL research addresses architecture, communication, and aggregation extensively, the operational lifecycle of FL systems remains underexplored in the research literature [10]. This paper contributes to closing that gap by examining how MLOps and FLOps practices can be combined into an architecture for production healthcare FL.

## 3 Methodology

This paper uses a structured literature review. The review includes foundational federated learning papers, healthcare FL research, privacy-preserving machine learning work, production ML literature, recent FLOps/FedOps studies, and healthcare governance sources. The analysis compares these sources across four dimensions: privacy protection, deployment scalability, operational complexity, and post-deployment governance.

The goal is to analyze how a federated learning system can be designed, deployed, monitored, and governed in a production healthcare environment.

## 4 Analysis and Findings

This section presents findings across the three research questions: FL deployment architecture (RQ1), privacy mechanism trade-offs (RQ2), and post-deployment governance practices (RQ3).

### 4.1 Containerized and Orchestrated FL Architecture

The first finding is that production-ready healthcare FL should be designed as a distributed software system, not only as a model-training method. In centralized ML, one team usually controls the data pipeline, training environment, model registry, and deployment target. In federated healthcare ML, those responsibilities are split across hospitals that may have different infrastructure, security policies, data schemas, approval workflows, and compute capacity. This makes standardization and orchestration necessary because the system must coordinate training without taking away local institutional control.

A practical architecture should separate the central control plane from the local hospital execution environment. The central control plane manages approved model code, container image versions, federation rounds, aggregation, model registry entries, and global audit records. Each hospital runs a local FL client in its own environment, trains on local data, validates the model locally, and sends only approved model updates or metrics back to the coordinator. This keeps raw patient data local while still allowing the federation to operate as one coordinated system. Figure 1 illustrates this reference architecture, showing the control-plane services, local hospital training and validation, and the flow from protected updates through secure aggregation and FedAvg to a candidate global model.

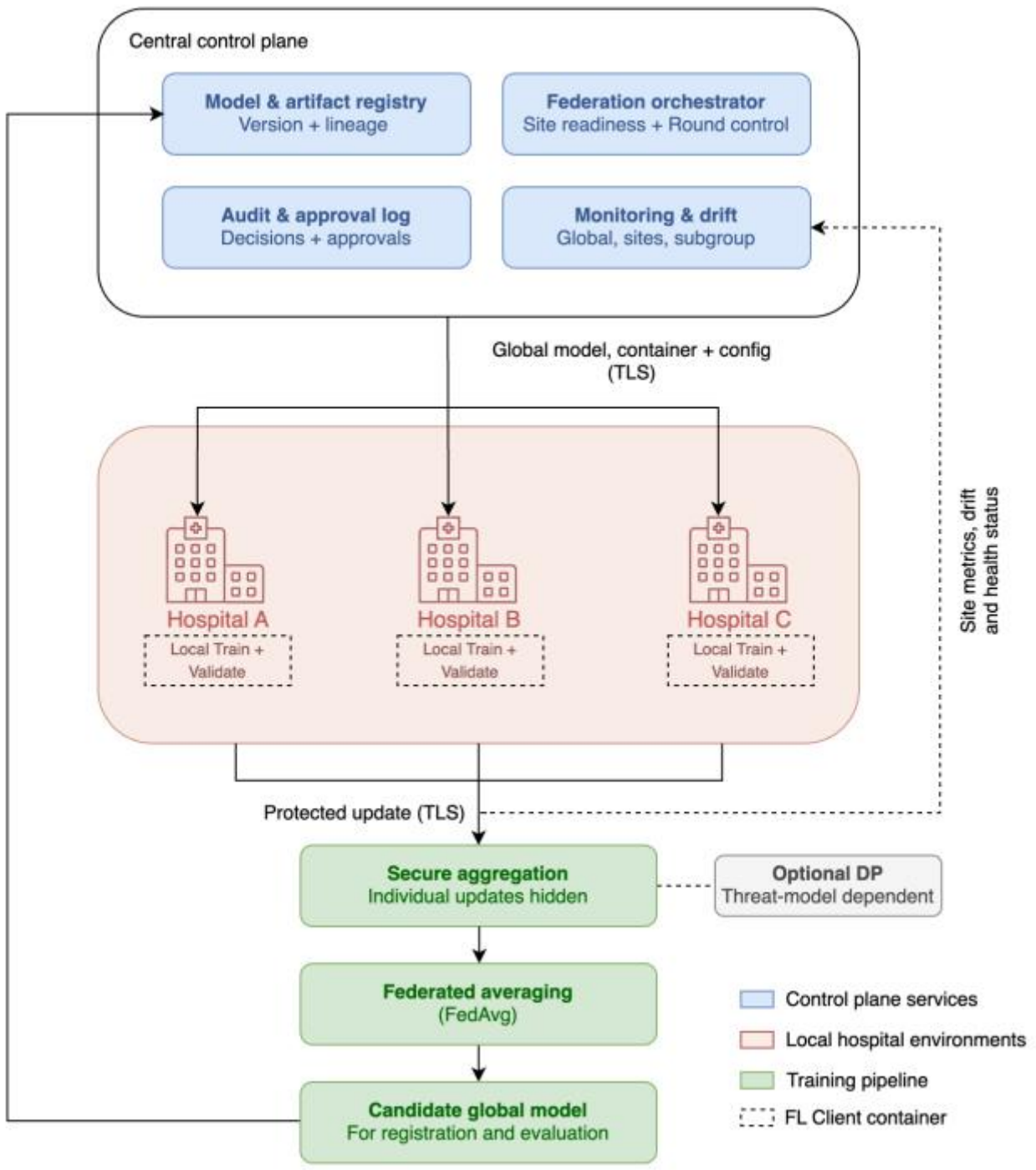

**Figure 1: Reference architecture for production-ready federated learning in healthcare, integrating central control-plane services, containerized local training, secure aggregation, and site-level monitoring.**

Containerization supports this architecture by packaging the training code, preprocessing logic, model dependencies, privacy tools, and logging components into a reproducible unit. This does not mean every site becomes identical, because hospitals may still have different data systems, hardware, and local configurations. However, it reduces one major source of variation: the software environment. If one hospital's model update behaves unexpectedly, the operations team can inspect the container version, code version, configuration, and logs instead of guessing which local dependency caused the issue.

Orchestration adds the operational layer that FL needs after packaging. It should manage deployment, health checks, client availability, communication retries, and controlled updates. This is where FedOps becomes useful. Moon et al. show that a FedOps platform can use Kubernetes and cloud-native components to support integration, testing, deployment, monitoring, and lifecycle management for FL systems [9]. For healthcare, this means orchestration should not only start training jobs. It should also decide which sites are ready to participate, which sites should be paused, when a training round has enough valid participants to continue, and when a failed round should be retried.

Non-IID data is also an important part of this architecture. In healthcare, each hospital may serve a different population. A pediatric hospital, rural clinic, and urban academic medical center may have different patient demographics, disease prevalence, equipment, and documentation practices. This can make local model updates behave differently across sites. If the system only tracks the global average, it may miss the fact that the model is improving for one site while getting worse for another. Therefore, the architecture should require per-site validation metrics, site-level drift checks, and participation metadata for every training round.

The key conclusion for RQ1 is that Docker/Kubernetes-style infrastructure is useful, but it is not the main goal by itself. The real goal is controlled, auditable, and repeatable cross-site learning. Containerization helps make local training more reproducible. Orchestration helps manage distributed execution. Governance decides when a model is safe enough to deploy.

## 4.2 Privacy, Utility, Scalability, and Complexity Trade-offs

The second finding is that privacy in healthcare FL must be layered. Keeping raw data local is necessary, but it is not enough. A hospital may never send patient records to the server, but model updates can still reveal information if an attacker can inspect or reconstruct them. Therefore, privacy should be treated as a system design problem, and not just as a feature of the FL algorithm.

Table 1 summarizes common privacy mechanisms for federated healthcare ML in terms of privacy role, model performance impact, and operational cost.

**Table 1: Privacy mechanisms for federated healthcare ML**

| Mechanism | Privacy role | Accuracy impact | Operational complexity |
|---|---|---|---|
| Secure aggregation | Hides individual updates | Low | Medium |
| Differential privacy | Limits record inference | Medium-High | Medium-High |
| Homomorphic encryption | Computes on encrypted data | Low | Very High |
| Encrypted communication | Protects data in transit | Low | Low-Medium |

Secure aggregation is often the most practical baseline for cross-silo healthcare FL. Its purpose is simple: the server should learn the combined update from participating hospitals, not each hospital's individual update [6]. This is useful because it reduces the risk that the coordinator can inspect a single institution's contribution. It also has less direct impact on model accuracy than differential privacy because it does not add noise to the update. The trade-off is operational. Secure aggregation adds key management, protocol coordination, and failure handling. If hospitals drop out during a round, the system needs clear rules for whether to continue, wait, or restart.

Differential privacy provides a stronger formal privacy guarantee by adding calibrated noise to limit what can be inferred about individual records [7]. This is valuable when the organization needs a measurable privacy bound. However, the cost is that too much noise can reduce model performance. In healthcare, this trade-off is serious as even a small performance drop may matter if the model supports clinical decision-making. Differential privacy also requires organizations to decide how much privacy loss is acceptable, continuously measure that loss during training, and establish rules for stopping or modifying training when the allowed limit is reached.

Encryption is also necessary, but it solves a different problem. Encrypting communication protects model updates while they move between hospitals and the coordinator. However, encryption in transit does not automatically protect the update after it is decrypted for aggregation. This is why encryption should be combined with secure aggregation, access control, and audit logging rather than treated as a complete privacy solution.

For RQ2, the practical implication is that no single mechanism is best for every healthcare FL deployment. A reasonable production design would use encrypted communication as a baseline, secure aggregation to reduce server visibility into individual updates, and differential privacy when formal privacy guarantees are required. The right combination depends on the threat model, clinical risk tolerance, infrastructure capacity, and regulatory expectations. Stronger privacy is valuable, but if it makes the system too slow, too inaccurate, or too difficult to operate safely, it can create a different kind of production risk.

## 4.3 Post-Deployment MLOps Governance

The third finding is that post-deployment governance is the main difference between a research FL prototype and a production healthcare FL system. In research, success is often measured by whether the model trains and reaches acceptable accuracy. In production healthcare, that is only the starting point. The system must remain reliable after deployment, even as hospitals change their data pipelines, patient populations, clinical protocols, and infrastructure. The candidate global model produced by the architecture in Figure 1 enters the governed lifecycle summarized in Figure 2: versioning and site-level validation precede staged rollout, monitoring results determine promotion or rollback, an audit trail records each decision, and post-deployment evidence can trigger federated retraining.

Model versioning must therefore include more than the global model file. Each federated round should record the model version, participating institutions, container image version, training configuration, privacy settings, aggregation method, per-site validation results, and approval status. This matters because failures in FL are not always caused by the model itself. A drop in performance could come from a changed preprocessing step at one hospital, a missing data field, a new patient population, or an unstable local training environment. Without lineage, it becomes difficult to know what changed and why the model behaved differently.

Data and transformation versioning are also necessary. In centralized ML, the training team may directly control the full data pipeline. In healthcare FL, each institution controls its own local data transformation logic. If the model input schema changes, every hospital must update its local preprocessing correctly. If one hospital maps a feature differently, the global model may receive inconsistent updates. This is why FLOps needs to track both the model version and the local data-interface version. The model is shared, but the data pipeline remains distributed.

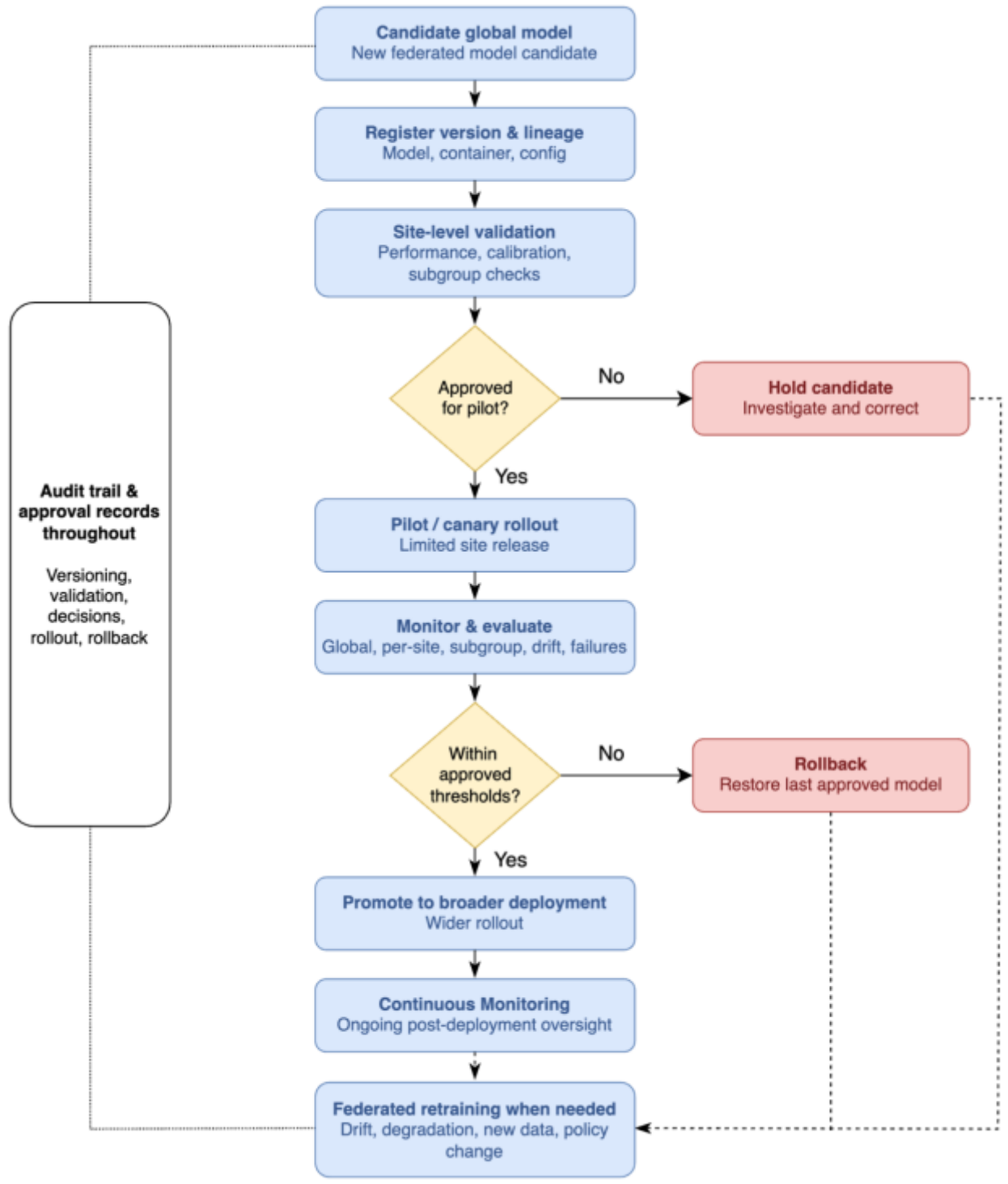

**Figure 2: Governed deployment and monitoring lifecycle for federated healthcare models, from version registration and site-level validation through staged rollout, rollback, continuous monitoring, and retraining.**

Monitoring must happen at both the global and local levels. A global metric can hide local failure. For example, the overall model may look stable while performance drops at a rural hospital, among elderly patients, or for a minority subgroup. Therefore, each site should report local validation metrics, drift indicators, calibration results, training failures, and communication status. The central coordinator should watch for sites that are diverging from the federation trend, while avoiding unnecessary exposure of patient-level data.

Client orchestration is also part of governance. A site should not automatically participate in every training round just because it is enrolled in the federation. The system should check whether the site is available, whether its software version is current, whether local validation passed, and whether its data pipeline is healthy. If a site repeatedly fails validation or sends abnormal updates, it should be paused until the issue is reviewed. This protects the global model from unreliable contributions and gives institutions a clear path to rejoin after fixing the issue.

Finally, production FL needs staged rollout and rollback. A new global model should first be validated locally, then released to a limited pilot group before broad deployment. If performance drops, drift increases, or a privacy control fails, the system should roll back to the last approved model version. Breck et al. describe canarying, monitoring, and rollback as important production ML practices [8]. In healthcare FL, these practices are not optional because model failure can affect patient care, institutional trust, and regulatory accountability.

Overall, production-ready healthcare FL needs a governance loop: version every artifact, validate at each site, monitor local and global behavior, audit every important decision, and roll back quickly when needed. This turns FL from a one-time collaborative training experiment into a sustainable clinical ML system.

## 5 Discussion

Federated learning shifts the production challenge rather than eliminating it. Instead of centralizing data, organizations must coordinate training, validation, privacy, monitoring, and governance across distributed institutions. This makes FL a strong fit for MLOps and FLOps analysis.

For RQ1, containerization and orchestration help make federated deployment reproducible and scalable, but they must be combined with security and governance controls. For RQ2, privacy mechanisms should be selected based on the specific threat model and operational constraints. Stronger privacy can increase complexity, reduce accuracy, or increase latency. For RQ3, the most important practices are model versioning, audit logging, drift monitoring, client orchestration, staged rollout, rollback, and clear accountability.

The practical recommendation is to design federated healthcare ML as a governed production system from the beginning. A privacy-preserving algorithm alone is not enough. The system also needs reliable deployment workflows, observable operations, documented decisions, and continuous monitoring. This is where the FLOps perspective is useful: it makes the lifecycle of federated systems explicit and shows why cross-silo FL needs more than a standard centralized MLOps pipeline.

## 6 Conclusion

Federated learning offers a promising path for collaborative healthcare AI because it reduces the need to centralize sensitive patient data. However, FL is not automatically private, reliable, or production-ready. Model updates may still leak information, and decentralized training creates new challenges for deployment, monitoring, debugging, reproducibility, and governance.

This paper argues that production-ready healthcare FL requires an integrated MLOps and FLOps architecture. Containerization and orchestration support reproducible deployment. Secure aggregation, differential privacy, and encryption provide different privacy trade-offs. Post-deployment practices such as model lineage, audit logging, drift monitoring, canary release, rollback, and governance are essential for long-term trust. In regulated healthcare environments, federated learning should be engineered not only for privacy, but also for reliability, accountability, and operational sustainability.

Future work should evaluate these trade-offs empirically in realistic cross-site healthcare deployments, especially under non-IID data, client dropout, privacy-budget constraints, and site-specific drift.